# Evolution of post-Accretion Induced Collapse Binaries: the Effect of Evaporation

Wei-Min Liu[1,2,3] and Xiang-Dong Li[1,3]

[1]*Department of Astronomy, Nanjing University, Nanjing 210046, China*

[2]*Department of Physics, Shangqiu Normal University, Shangqiu 476000, China*

[3]*Key Laboratory of Modern Astronomy and Astrophysics (Nanjing University), Ministry of Education, Nanjing 210046, China*

*lixd@nju.edu.cn*

## ABSTRACT

Accretion-induced collapse (AIC) is widely accepted to be one of the formation channels for millisecond pulsars (MSPs). Since the MSPs have high spin-down luminosities, they can immediately start to evaporate their companion stars after birth. In this paper we present a detailed investigation on the evolution of the post-AIC binaries, taking into account the effect of evaporation both before and during the Roche-lobe overflow process. We discuss the possible influence of the input parameters including the evaporation efficiency, the initial spin period and the initial surface magnetic field of the newborn neutron star. We compare the calculated results with the traditional low-mass X-ray binary evolution, and suggest that they may reproduce at least part of the observed redbacks and black widows in the companion mass - orbital period plane depending on the mechanisms of angular momentum loss associated with evaporation.

*Subject headings:* stars: evolution - stars: white dwarfs - stars: pulsars - pulsars: radio - X-rays: binaries

## 1. INTRODUCTION

Millisecond pulsars (MSPs) are characterized by short spin-periods ($P \lesssim 30$ ms), weak surface magnetic fields ($B \sim 10^8 - 10^9$ G), and low spin-down rates ($\dot{P} \sim 10^{-19} - 10^{-21}$ s s$^{-1}$). Currently, there are two scenarios for the formation of MSPs (Bhattacharya & van den Heuvel 1991, for a review). One is the recycling model which involves mass transfer in low-mass X-ray binary (LMXBs) containing a neutron star (NS). By accreting mass and angular momentum (AM) from the donor star, the NS can be spun up to spin periods of order of milliseconds with its magnetic field significantly decayed (Alpar et al. 1982). In the other model, MSPs can be formed by accretion-induced collapse (AIC) of massive white dwarfs (WDs). When an ONeMg WD grows its mass to the Chandrasekhar mass limit ($\sim$1.38 $M_\odot$) by accreting material from the companion star, the electron capture process will lead to collapse of the WD and result in a low-field, rapidly rotating NS (Nomoto & Kondo 1991).

Once a MSP is formed, it will start to evaporate its companion with its high-energy radiation/particles if there is no mass transfer interaction between them (van den Heuvel & van Paradijs 1988; Ruderman et al. 1989). This evaporation process has been invoked to explain the two sub-populations of MSPs called black widows and redbacks which show regular radio eclipses around superior conjunction (Fruchter et al. 1988; Stappers et al. 1996; Roberts 2013, and references therein). Based on the recycling scenario, several models have been explored to connect the evolution of LMXBs with the formation of redbacks and black widows, in which temporary termination of mass accretion is assumed to be caused by disrupted magnetic braking (MB) (Chen et al. 2013), irradiation-induced cyclic mass transfer (Benvenuto et al. 2014), and thermal-viscous instability in the accretion disks (Jia & Li 2015). In a subsequent work, Jia & Li (2016) systemically investigated the impact of the evaporation on the evolution of LMXBs by considering different angular momentum loss (AML) mechanisms and evaporation efficiency. Their calculations show that systems with initially tight orbits possibly experience dynamically unstable mass transfer and lead to the formation of isolated pulsars.

Based on binary population synthesis (BPS) calculations, Hurley et al. (2010) emphasized the contribution of the AIC channel to the formation of MSPs (see however Zhu et al. 2015). Detailed calculations on the formation processes of MSPs via AIC have been carried out by some authors (Sutantyo & Li 2000; Ivanova & Taam 2004; Tauris et al. 2013). A MSP produced by the AIC events should immediately start to evaporate its companion star, since mass transfer is terminated after AIC because of the sudden orbital expansion due to gravitational mass loss. However, few works consider the effect of evaporation on the subsequent evolution. Smedley et al. (2015) investigated the formation of redbacks based on the post-AIC evolution with energetic evaporation by the pulsars taken into account, but their study was limited to systems consisting of a $1.2 M_\odot$ WD + a $1.0 M_\odot$ donor star in a 0.3 day orbit. Ablimit & Li (2015) considered the formation of MSPs by a particular AIC channel under self-excited wind-driven evolution, and briefly discussed the influence of ablation of the companion star on the possible formation of redbacks. Obviously a systematic study on this subject is needed.

In this paper, we carry out calculations on the evaporation process of MSPs born through AIC. We focus on the evaporation effect and the possible link between the post-AIC binaries with redbacks and black widows. In our calculations, different choices of the initial parameters such as the spin periods and the surface magnetic fields of the newborn NSs, the orbital periods, and the evaporation efficiencies are taken into account. The rest of this paper is organized as follows. We give a detailed description of our model in Section 2. The calculated results are presented in Section 3. Our discussion and conclusions are in Section 4.

## 2. MODEL

We calculate the binary evolution using Modules for Experiments in Stellar Astrophysics (MESA) (Paxton et al. 2011, 2013, 2015). The whole evolution can be divided into two stages: (1) the pre-AIC stage, in which a massive ONeMg WD accretes from its companion star until its mass reaches $1.38M_\odot$ and the WD collapses to form a NS; (2) the post-AIC stage, in which the binary continues to evolve with evaporation, mass transfer and mass loss.

### 2.1. The pre-AIC stage

In this stage, the progenitor binaries are composed by a WD (defined as the primary, of mass $M_{\rm WD}$) and a main sequence (MS) donor star (of mass $M_{\rm d}$). We take Solar chemical abundance ($X = 0.70$, $Y = 0.28$, and $Z = 0.02$) for the donor star. Its Roche-lobe (RL) radius is evaluated with the Eggleton (1983) formula

$$\frac{R_{\rm L,d}}{a} = \frac{0.49q^{2/3}}{0.6q^{2/3} + \ln(1 + q^{1/3})}, \tag{1}$$

where $q = M_{\rm d}/M_{\rm WD}$ is the mass ratio and $a$ is the orbital separation. We adopt the Ritter (1988) scheme in MESA to calculate the mass transfer rate via Roche-lobe overflow (RLOF). The growth of the mass of an accreting ONeMg WD depends on how much mass can be accumulated during the hydrogen and helium burning processes. The mass growth rate $\dot{M}_{\rm WD}$ of a WD can be described as follows,

$$\dot{M}_{\rm WD} = -\eta_{\rm H}\eta_{\rm He}\dot{M}_{\rm d}, \tag{2}$$

where $\eta_{\rm H}$ and $\eta_{\rm He}$ are the accumulation efficiencies for hydrogen and helium burning respectively, and $-\dot{M}_{\rm d}$ is the mass transfer rate. We adopt the numerically calculated results by Hillman et al. (2016) for $\eta_{\rm H}$, and the prescriptions suggested by Kato & Hachisu (2004) for $\eta_{\mathrm{He}}$ (see also Liu & Li 2016). We further assume that the rest matter is ejected from the system in the form of isotropic wind at a rate of $(|\dot{M}_{\mathrm{d}}| - \dot{M}_{\mathrm{WD}})$, carrying the specific AM of the WD (Hachisu et al. 1996).

AML due to gravitational radiation (GR) and MB is included in our calculations. The AML rate caused by GR is given by (Landau & Lifshitz 1975),

$$\dot{J}_{\mathrm{GR}} = -\frac{32G^{7/2}}{5c^5} \frac{M_{\mathrm{WD}}^2 M_{\mathrm{d}}^2 (M_{\mathrm{WD}} + M_{\mathrm{d}})^{1/2}}{a^{7/2}} \ , \tag{3}$$

where $c$ is the speed of light and $G$ is the gravitational constant. We assume that MB is confined to MS stars with mass in the range of $\sim 0.3 - 1.5\,M_\odot$, and adopt the default setting in MESA for the AML rate (i.e., the $\gamma = 3$ case for Eq. (36) in Rappaport et al. 1983),

$$\dot{J}_{\mathrm{MB}} = -3.8 \times 10^{-30} M_{\mathrm{d}} R_{\mathrm{d}}^3 R_\odot \omega^3 \ \mathrm{dyn\,cm}, \tag{4}$$

where $R_{\mathrm{d}}$, $R_\odot$, and $\omega$ are the radii of the donor star and the Sun, and the angular velocity of the binary, respectively.

## 2.2. AIC and the post-AIC stage

During the pre-AIC stage, if the mass of the WD reaches $1.38\,M_\odot$, AIC of the WD is assumed to take place. The WD thus collapses into a NS of mass $M_{\mathrm{NS}} = 1.25\,M_\odot$, and the excess mass $(0.13\,M_\odot)$ is converted to the NS's binding energy. The binary orbit accordingly changes with (Verbunt et al. 1990),

$$\frac{a_{\mathrm{circ}}}{a_{\mathrm{pre}}} = \frac{M_{\mathrm{WD}} + M_{\mathrm{d}}}{M_{\mathrm{NS}} + M_{\mathrm{d}}}, \tag{5}$$

where $a_{\mathrm{pre}}$ and $a_{\mathrm{circ}}$ are the pre-AIC orbital separation and the post-AIC orbital separation after tidal circularization, respectively[1]. This immediately ceases the mass transfer because the sudden expansion of the binary orbit makes the companion to be detached from its RL. The newborn NS acts as a rapidly rotating radio pulsar and starts to evaporate its companion. The related wind mass loss rate can be expressed as (van den Heuvel & van Paradijs 1988),

$$\dot{M}_{\mathrm{d,\,evap}} = -\frac{f}{2v_{\mathrm{d,\,esc}}^2} L_{\mathrm{p}} \left(\frac{R_{\mathrm{d}}}{a}\right)^2 , \tag{6}$$

---

[1] The binary orbit becomes eccentric because of mass loss during AIC. Here we assume that tides can efficiently circularize the orbit.

where $v_{\rm d,esc}$ is the surface escape velocity of the donor star, $L_{\rm p} = 4\pi^2 I\dot{P}/P^3$ is the spin-down luminosity of the pulsar, and $f$ is the efficiency factor of evaporation. Here $I$, $P$, and $\dot{P}$ are the moment of inertia, the spin period of the pulsar and its derivative, respectively. We assume that the evaporative material takes away the specific AM of the companion (alternative prescription for AML will be discussed later), and that the NS spin evolution during the radio pulsar phase is described by the standard magnetic dipole radiation model (Shapiro & Teukolsky 1983).

With nuclear expansion of the companion or orbital shrinkage due to MB and mass loss, mass transfer via RLOF will resume. An accretion disk is then formed around the NS. We first discriminate whether the disk is subject to thermal-viscous instability. If the mass transfer rate is higher than a critical value given by (Dubus et al. 1999),

$$\dot{M}_{\rm cr} \simeq 3.2 \times 10^{-9} \left(\frac{M_{\rm NS}}{1.4M_\odot}\right)^{0.5} \left(\frac{M_{\rm d}}{1.0M_\odot}\right)^{-0.2} \left(\frac{P_{\rm orb}}{1.0\,{\rm d}}\right)^{1.4} M_\odot {\rm yr}^{-1}, \tag{7}$$

where $P_{\rm orb}$ is the orbital period, the accretion disk is stable, and in this case we define $\beta$ as the fraction of the accreted matter by the NS to the lost matter from the companion via RLOF. In our calculation, we fix its value to be 0.5, and the excess matter is assumed to be expelled by the NS in the isotropic wind carrying the specific AM of the NS. Meanwhile, the accretion rate of the NS is limited by the Eddington accretion rate $\dot{M}_{\rm Edd}$, so the actual accretion rate of the NS is

$$\dot{M}_{\rm NS} = \min(-\beta\dot{M}_{\rm d}, \dot{M}_{\rm Edd}), \tag{8}$$

where

$$\dot{M}_{\rm Edd} = 3.6 \times 10^{-8} \left(\frac{M_{\rm NS}}{1.4\,M_\odot}\right) \left(\frac{0.1}{GM_{\rm NS}/R_{\rm NS}c^2}\right) \left(\frac{1.7}{1+X}\right) M_\odot {\rm yr}^{-1}. \tag{9}$$

Here $X$ is the hydrogen abundance of the accreting matter, and $R_{\rm NS}$ is the NS radius (taken to be $10^6$ cm). If $-\dot{M}_{\rm d} < \dot{M}_{\rm cr}$, the accretion disk becomes thermally and viscously unstable, and the binary alternates between short outbursts and long quiescence. Same as in Jia & Li (2015) we assume that the evaporation process commences and the NS does not accretes any matter from the companion in this episode.

Now we can list the total AML rate during the post-AIC evolution as follows,

$$\dot{J} = \dot{J}_{\rm GR} + \dot{J}_{\rm MB} + \dot{J}_{\rm ML,NS} + \dot{J}_{\rm ML,evap}. \tag{10}$$

The first and second terms on the right hand side of Eq. (10) are due to GR and MB, respectively. The third and fourth terms correspond to AML caused by mass loss from the NS and evaporative wind loss from the companion star, and we assume that the lost matter carries the specific AM from the NS and from the surface of the companion star, respectively (see also Jia & Li 2016).

The orbital AM of the binary system is

$$J = \left(\frac{G}{\sqrt{2\pi}}\right)^{2/3} M_{NS} M_d \left(\frac{P_{orb}}{M}\right)^{1/3}, \tag{11}$$

where $M$ is the total mass of the binary. Differentiation of the above equation with time gives

$$\frac{\dot{J}}{J} = \frac{\dot{M}_{NS}}{M_{NS}} + \frac{\dot{M}_d}{M_d} + \frac{1}{3}\frac{\dot{P}_{orb}}{P_{orb}} - \frac{1}{3}\frac{\dot{M}}{M}. \tag{12}$$

Therefore, we can derive the orbital evolution in three different cases.

(1) In the pre-RLOF phase, $\dot{M}_{NS} = 0$, $\dot{M} = \dot{M}_{d,evap}$, where $\dot{M}_{d,evap}$ is the mass loss rate of the companion star by evaporation. Combining Eqs. (10) and (12) we obtain

$$\frac{\dot{P}_{orb}}{P_{orb}} = \frac{3(\dot{J}_{GR} + \dot{J}_{MB} + \dot{J}_{ML,evap})}{J} - \left(\frac{3M - M_d}{M M_d}\right)\dot{M}_{d,evap}. \tag{13}$$

(2) In the stable disk accretion phase, $\dot{M}_{d,evap} = 0$. Here we use $\dot{M}_{d,RLOF}$ to denote the mass loss rate from the donor star via RLOF. If $-\dot{M}_{d,RLOF} < 2.0 \times \dot{M}_{Edd}$, then

$$\frac{\dot{P}_{orb}}{P_{orb}} = \frac{3(\dot{J}_{GR} + \dot{J}_{MB} + \dot{J}_{ML,NS})}{J} - 3\left(\frac{1}{M_d} - \frac{1}{6M} - \frac{1}{2M_{NS}}\right)\dot{M}_{d,RLOF}. \tag{14}$$

Otherwise we have

$$\frac{\dot{P}_{orb}}{P_{orb}} = \frac{3(\dot{J}_{GR} + \dot{J}_{MB} + \dot{J}_{ML,NS})}{J} - 3\left(\frac{1}{M_{NS}} - \frac{1}{3M}\right)\dot{M}_{Edd} - 3\left(\frac{1}{M_d} - \frac{1}{3M}\right)\dot{M}_{d,RLOF}. \tag{15}$$

(3) In the unstable disk accretion phase there are both evaporative wind loss and RLOF mass transfer but no accretion, i.e., $\dot{M}_{NS} = 0$, $\dot{M} = \dot{M}_{d,evap} + \dot{M}_{d,RLOF}$. We then obtain

$$\frac{\dot{P}_{orb}}{P_{orb}} = \frac{3(\dot{J}_{GR} + \dot{J}_{MB} + \dot{J}_{ML,NS} + \dot{J}_{ML,evap})}{J} - \left(\frac{3M - M_d}{M M_d}\right)(\dot{M}_{d,evap} + \dot{M}_{d,RLOF}). \tag{16}$$

In the following we demonstrate the calculated binary evolution in different cases.

## 3. RESULTS

Figure 1 shows the distribution of the incipient WD binary systems in the donor mass ($M_d$) - orbital period ($P_{orb}$) plane. Here we fix the initial WD mass to be $1.2\,M_\odot$. The initial donor star mass and the initial orbital period are set to be in the range of $2.0 - 3.5 M_\odot$ and $0.4 - 5.0$ days, respectively. The Figure can be divided into four regions: in the region filled with crosses in squares the donor star overflows its RL at the beginning of the calculation; in the region with crosses the binaries will experience (delayed) dynamically unstable mass transfer and possibly lead to common envelope (CE) evolution; in the region with open circles the WDs undergo nova eruptions during the mass transfer process and cannot grow their masses to $1.38\,M_\odot$ within a Hubble time; only in the region with solid circles can AIC occur. This result is largely consistent with previous works (Li & van den Heuvel 1997; Tauris et al. 2013)[2].

Figure 2 compares the distribution of the binaries before (denoted by blue open circles) and after (denoted by red symbols) AIC in the $M_d - P_{orb}$ plane. Only the orbital period is changed during AIC, and lower donor mass leads to larger change in the orbital period. In Fig. 2, we use different types of red symbols to denote the possible post-AIC evolutionary results: the circles, crosses and open squares represent that the systems will experience stable mass transfer, dynamically unstable mass transfer, and no RLOF within a Hubble time, respectively. We can find that the subsequent evolution depends on the values of $f$ since the evaporative winds changes the donor mass and orbital period. With increasing $f$, fewer systems will undergo unstable mass transfer because the donor mass is smaller at the beginning of RLOF; meanwhile more systems will remain detached during the evolution. The NS binaries that can proceed stable mass transfer have orbital periods between 0.5 and 2.0 days and donor masses between 0.9 and 2.2 $M_\odot$ approximately. It is interesting to see that these distributions are relatively narrow compared with those in the standard recycling model (e.g., Liu & Chen 2011).

Once a NS is born, it starts to evaporate the companion with part of its rotational energy. Here we assume that the initial spin-period ($P_i$) and the surface magnetic field ($B_{surf,i}$) of the NS are 3 ms and $3.0 \times 10^8$ G, respectively. Below we will consider other choices of the spin periods and magnetic fields. Since the magnetic field of the NS does not decay in the radio pulsar stage, we can deduce the derivative of the spin period according to the following formula

$$\dot{P} = 10^{-39}\frac{B^2}{P}. \tag{17}$$

Figure 3 displays the distribution of the binaries in the $P_{orb} - M_d$ plane at the onset of the (second) RLOF mass transfer. At this moment the NS starts to accrete and evaporation stops. Here we show three cases with $f = 0.02$ (top), 0.1 (middle), and 0.3 (bottom). When $f$ increases, the donor mass becomes lower, and the binaries evolve into a bimodal configuration, which results from the competition between mass loss and MB (see Eq. [13]). In our prescription for AML, the evaporative winds always widen the orbit, while MB always shrinks the orbit. The red and blue solid circles correspond to two evolutionary examples that will be displayed, with the initial orbital period and donor star mass of 0.775 day and 1.406 $M_\odot$, and 1.305 days and 1.981 $M_\odot$, respectively.

---

[2]We do not consider the binary evolution with a red-giant donor since their contribution to AIC is rather small.

Figure 4 shows the distribution of the binaries at end of the evolution with $f = 0$, 0.02, 0.1, and 0.3, corresponding to either the donor star having evolved to be a WD or the age of the binary exceeding Hubble time. The binaries with orbital periods at the beginning of the second RLOF above the so-called bifurcation period (Pylyser & Savonije 1988, 1989) evolve into low-mass binary pulsars (LMBPs) with a helium WD companion (depicted with circles), while those with orbital periods below the bifurcation period evolve into compact LMXBs with sub-stellar companion stars (depicted with triangles). To see how the bifurcation period changes with mass transfer and evaporation, in Fig. 5 we present the orbital period evolution for five NS binaries with different $f$ values. It seems that the bifurcation periods are quite close to each other ($\sim 1$ day), and increase slightly with increasing $f$.

The solid line in the four panels in Fig. 4 is the theoretical $M_{WD} - P_{orb}$ relation for LMBPs given by Tauris & Savonije (1999). We can see that if there is no evaporation in the post-AIC evolution, the resulting distribution is well consistent with the theoretical relation, same as in LMXB evolution. However, once evaporation is included, the distribution deviate from the theoretical prediction, especially in the case of $f = 0.3$: for a given $M_{WD}$, the orbital period is longer than expected. The reason is that in the late stage of the evolution, the orbital evolution and mass loss are mainly driven by evaporation rather than mass transfer via RLOF (see also Jia & Li 2016).

To demonstrate the detailed evolutionary processes, in Figs. 6 and 7 we present two representative examples for the post-AIC binaries which are marked in Figs. 3 and 4. Figures 6 and 7 correspond to the blue circle with $P_{orb,i} = 1.305$ days and $M_{d,i} = 1.981\,M_\odot$ and the red triangle with $P_{orb,i} = 0.775$ day and $M_{d,i} = 1.406\,M_\odot$, respectively. Figure 8 compares the rates of AML caused by different mechanisms in the evolutionary paths presented in Figs. 6 (left panel) and 7 (right panel). In Figs. 6 and 7, the left panel displays the evolution of the mass transfer rate (black solid line), the evaporative wind loss rate (red line), and the donor mass (blue line). The black dashed line represents the critical mass transfer rate for disk instability. The right panel depicts the evolution of the NS mass (solid line) and the orbital period (dotted line). The values of $f$ are 0, 0.02, 0.1, and 0.3 from top to bottom. In Fig. 6, when $f = 0$, the evolution is very similar to standard LMXB evolution driven by nuclear evolution of the donor star (Tauris & Savonije 1999). In the other cases, evaporation takes place before the occurrence of RLOF and when the accretion disk becomes unstable. With increasing $f$, the donor mass becomes smaller at the onset of RLOF, and the mass transfer rate decreases. AML associated with evaporative winds dominates mass loss and the orbital evolution (also see the left panel of Fig. 8 for a comparison of the AML rates), and this causes the final orbital period to be smaller than without evaporation. Figure 7 presents the cataclysmic variable (CV)-like evolution for compact LMXBs. A two-stage evaporation is clearly seen: one is during the detached phase after AIC, the other starts when the MB effect stops and the binary enters in the period gap. From then on the mass transfer rate is always less than the critical value for disk instability, so evaporation continues until Hubble age. The evolutionary tracks with different values of $f$ seem to be similar. The reason is that in late evolution AML is dominated by GR rather than evaporation in this case (also see the right panel of Fig. 8 for a comparison of the AML rates).

Figure 9 compares some selected evolutionary tracks with known redbacks (orange squares) and black widows (green squares)[3]. In the left and right panels $f = 0.1$ and 0.3, respectively. In the upper and lower panels, we assume that the evaporative winds carry the specific AM from the surface of the donor star or from the inner Lagrangian point (corresponding to Mode A and B winds in Jia & Li 2016), respectively. The blue lines denote evaporation during the RL decoupling, the black lines stable disk accretion, and the red lines unstable disk accretion with evaporation. We can see that in the case of $f = 0.3$, Mode A, a large fraction of the redbacks can be covered by the evolutionary paths. In this case, due to efficient evaporation following the AIC event, the donor masses become relatively small at the onset of the second RLOF compared with in the cases of $f = 0.1$, Modes A and B (see, e.g., Fig. 3). Therefore, most of the binaries with $P_{orb} \sim 0.1 - 1$ day at the onset of the second RLOF will evolve to pass the redback zone. In the case of $f = 0.3$, Mode B, because AML associated with the first evaporation is less efficient, the binaries tend to evolve to wider orbits and it is less promising for them to produce redbacks. The formation of black widows requires that the orbital periods at the onset of RLOF are distributed within a narrow range around 0.5 day (e.g., Benvenuto et al. 2014; Jia & Li 2016). A small fraction of the binaries in our cases can finally evolve to be black widows (with $P_{orb} < 0.1$ day), suggesting that most black widows may originate from normal LMXBs rather the post-AIC binaries.

In the following we discuss the possible influence of the input parameters. The spin periods of the newborn NS depends on the spin AM of the WD just under collapse, which is difficult to determine because it is related to the magnetic field strength and the accretion rate of the WD. In Figs. 10 and 11, we show the evolutionary tracks with same initial conditions as in Figs. 6 and 7 respectively, but with $P_i = 10$ ms for the NS. According to magnetic dipole radiation theory this reduces the spin-down luminosity by roughly two orders of magnitude, so the evolutions do not have remarkable difference from those without evaporation, although evaporation still proceeds.

---

[3]Data are taken from A. Patruno's MSP catalogue, https://apatruno.wordpress.com/about/millisecond-pulsarcatalogue/

Although it is usually assumed that NSs formed by AIC have weak magnetic fields, there is no evidence that their initial magnetic fields cannot be strong. Figures 12 and 13 are similar to Figs. 6 and 7 respectively, but with $B_{surf,i} = 1.0 \times 10^{12}$ G. The solid and dashed lines correspond to $f = 0$ and 0.3, respectively. Because of the strong magnetic field, the NS rapidly spin-down during the radio pulsar stage. In Fig. 12 the NS evolves beyond the death-line and ceases evaporation even before the onset of RLOF. Thus evaporation plays a minor role in the evolution for strong-field NSs.

## 4. DISCUSSION AND CONCLUSIONS

In this work we study the evolution of NSs formed by AIC, paying particular attention to possible evaporation of its companion star. These NSs are usually regarded as a sub-population of MSPs, together with those recycled in LMXBs. The AIC model has been extensively investigated, but there are few works that discuss the effect of evaporation of the donor stars. Smedley et al. (2015) explored possible formation of redbacks in the AIC model. For the progenitor binaries they adopted initial parameters with $1.2\,M_\odot$ WD, $1.0\,M_\odot$ donor, and 0.3 day orbital period. According to our Fig. 1 (see also Tauris et al. 2013), it is very difficult for the WD in such a binary to reach a mass of $1.38\,M_\odot$ and lead to AIC, unless some unusual mechanisms such as irradiation-excited winds from the donor are invoked to accelerate the mass transfer (Ablimit & Li 2015).

Evaporation is prolonged in the AIC model compared with in the traditional recycling model, so its effect may be more prominent. In LMXBs evaporation takes place only when the following two conditions are satisfied: (1) the NS has been spun up to periods of milliseconds by accreting sufficient mass (typically $0.1\,M_\odot$), and (2) the accretion rate onto the NS drops significantly, so that the accreting NS transits into a MSP. Indeed the three so-called transitional pulsars are all redbacks (e.g., Archibald et al. 2009; Papitto et al. 2013; de Martino et al. 2013). In the AIC model we assume that the newborn NS appears as a MSP and starts evaporation immediately, because (1) AM conservation during the collapse implies that the NS should be rapidly rotating, and (2) mass loss during the collapse induces a wider orbital separation and stops the mass transfer. Thus, depending on the spin-down luminosity and the efficiency of evaporation, the mass of the companion star can be reduced before the second RLOF, although the subsequent evolution seems to be similar to that of

LMXBs.

Our calculated results for the post-AIC evolution can be summarized as follows.

(1) If the NS is born as a MSP, evaporation can considerably decrease the mass of the companion star. This influences the parameter space of the binaries for subsequent successful mass transfer.

(2) After the second RLOF starts, the binaries evolve like LMXBs: those with orbital periods above the bifurcation period evolve to be MSP+He WD binaries, and those below the bifurcation period evolve along the CV-like evolutionary tracks. In the former case, the distribution of the MSP+He WD binaries deviates from the standard $M_{WD} - P_{orb}$ relation, with longer $P_{orb}$ for a given $M_{WD}$ because mass loss and AML are dominated by the evaporative winds. In the latter case evaporation plays a minor role in AML as compared with GR and MB.

(3) The evolutionary paths, if considering the possible range of the input parameters, can cover the distribution of part of redbacks and black widows, depending on the AML mechanisms of evaporative winds.

(4) Finally, if the NS is born with relatively longer spin period or stronger magnetic field, its spin-down luminosity is either too low or decreases rapidly, so the effect of evaporation can be largely neglected.

We are grateful to the referee for helpful comments. This work was supported by the National Key Research and Development Program of China (2016YFA0400803), the Natural Science Foundation of China under grant Nos. 11133001, 11333004 and 11573016, and the Program for Innovative Research Team (in Science and Technology) at the University of Henan Province.

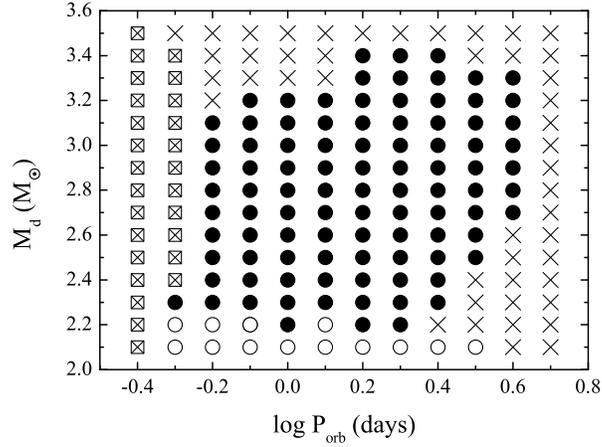

Fig. 1.— Different symbols in the initial orbital period vs. the companion mass diagram indicate different evolutionary fate of the WD binaries: crosses - systems will experience dynamically unstable mass transfer; crosses in squares - the forbidden region where the companion's radius exceeds its RL radius at the beginning; open circles - the WDs will experience nova eruptions; solid circles - the WDs can accrete matter to reach $1.38\,M_\odot$ and trigger AIC. The initial WD mass is taken to be $M_{\rm WD} = 1.2\,M_\odot$.

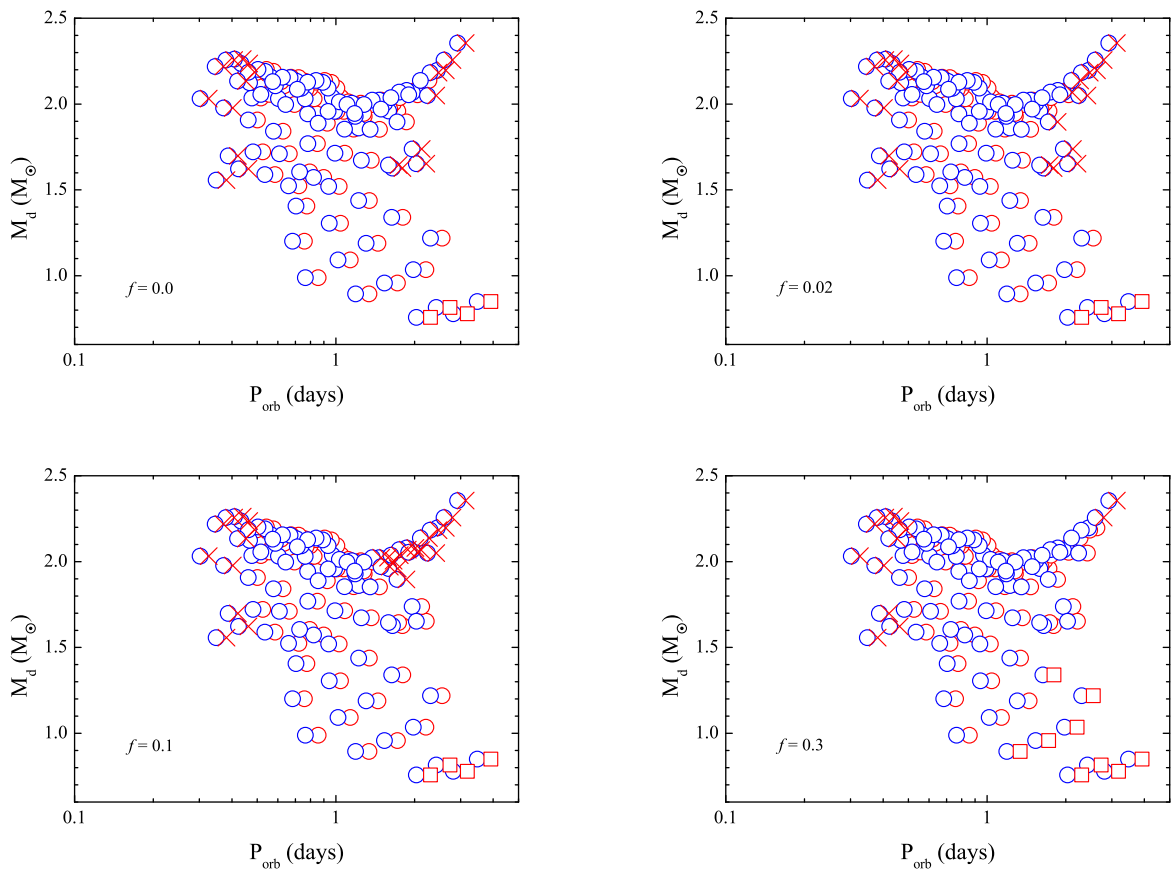

Fig. 2.— The blue circles and the red symbols represent the distribution of the binaries just before and after AIC, respectively. Red circles: successful evolution; red crosses: mass transfer will be dynamically unstable; red squares: cannot overflow the RL within a Hubble time.

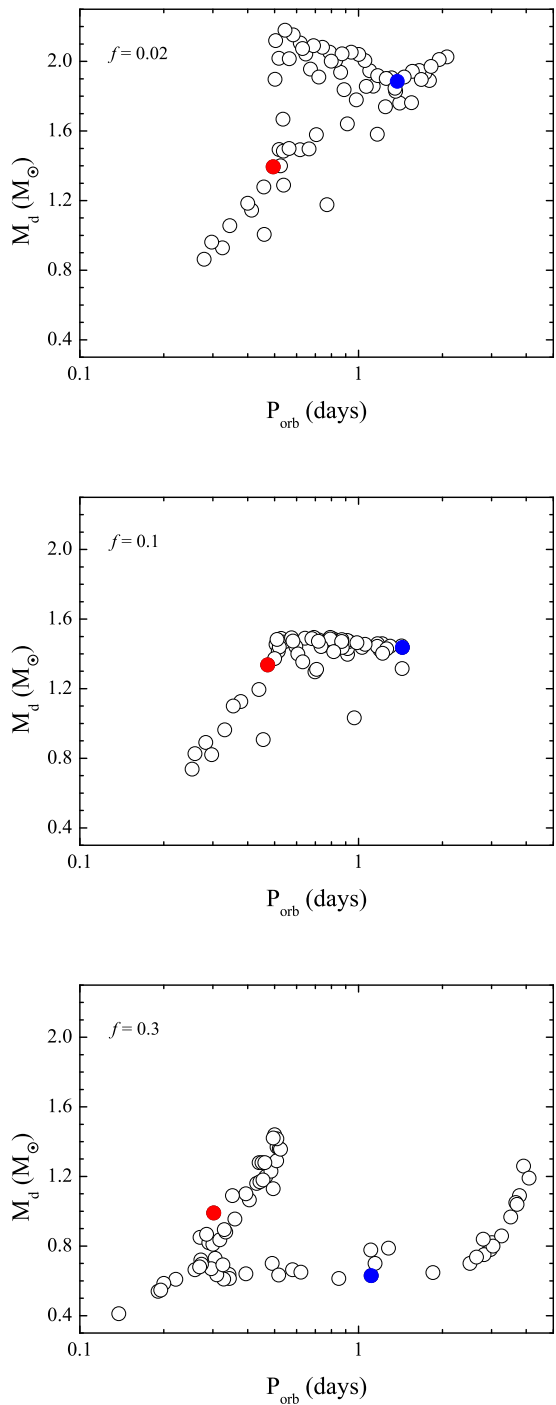

Fig. 3.— Distribution of the binaries in the orbital period vs. donor mass diagram at the onset of second RLOF for different values of $f$. The initial parameters for the NS are $P_i = 3$ ms and $B_{surf,i} = 3.0 \times 10^8$ G. The blue and red solid circles denote the current positions of two example binaries with initial parameters $P_{orb,i} = 1.305$ days and $M_{d,i} = 1.981 \, M_\odot$, and $P_{orb,i} = 0.775$ day and $M_{d,i} = 1.406 \, M_\odot$, respectively.

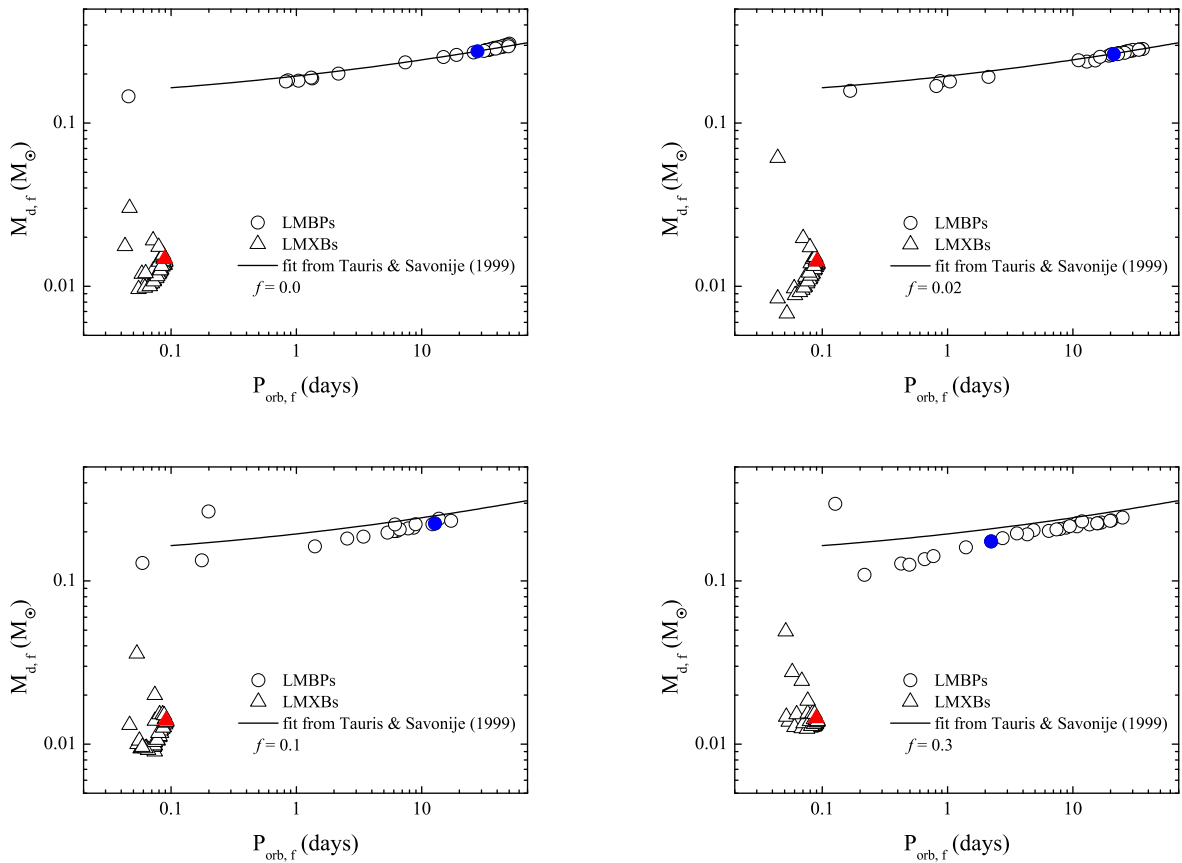

Fig. 4.— Distribution of the final orbital period and donor mass under different values of $f$. The blue and red symbol have the same meanings as in Fig. 2. The solid line is the theoretical $P_{\mathrm{orb}}$ - $M_{\mathrm{d}}$ relation of LMBPs given by Tauris & Savonije (1999).

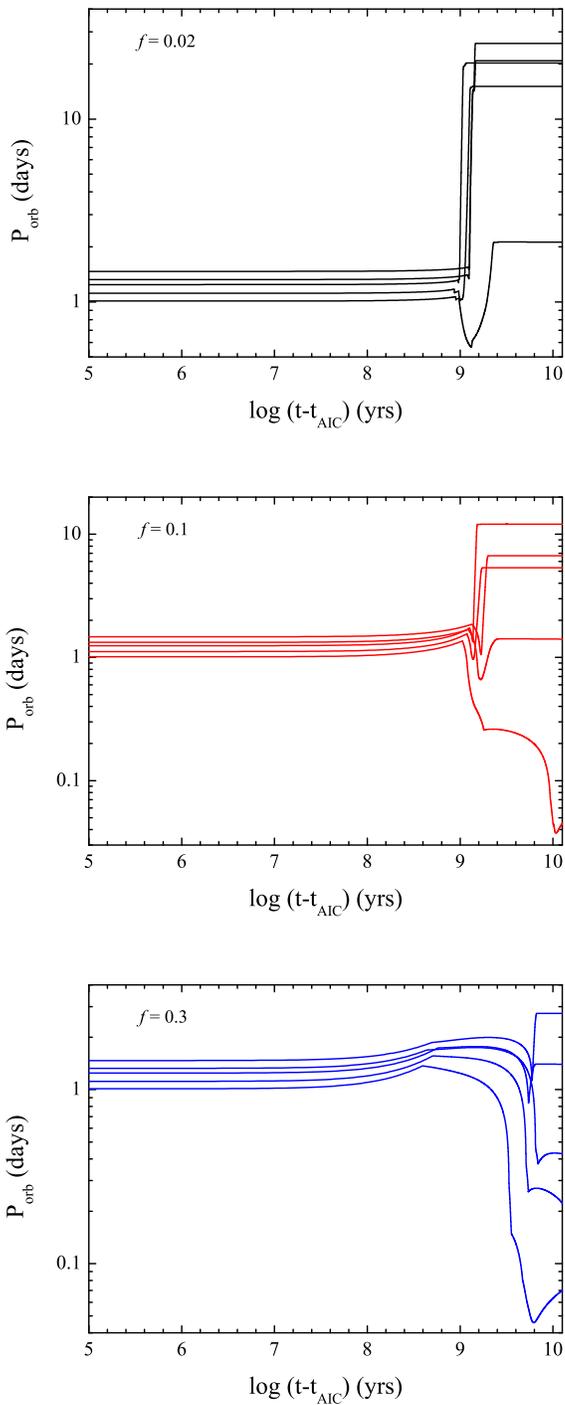

Fig. 5.— Post-AIC evolution of the orbital period under different values of $f$ for five NS binaries. Here $t_{AIC}$ denotes the time at AIC. From top to bottom the initial parameters are: $P_{orb,i} = 1.469$ days and $M_{d,i} = 1.853\,M_\odot$, $P_{orb,i} = 1.325$ days and $M_{d,i} = 1.861\,M_\odot$, $P_{orb,i} = 1.244$ days and $M_{d,i} = 2.005\,M_\odot$, $P_{orb,i} = 1.115$ days and $M_{d,i} = 2.013\,M_\odot$, and $P_{orb,i} = 1.011$ days and $M_{d,i} = 1.957\,M_\odot$.

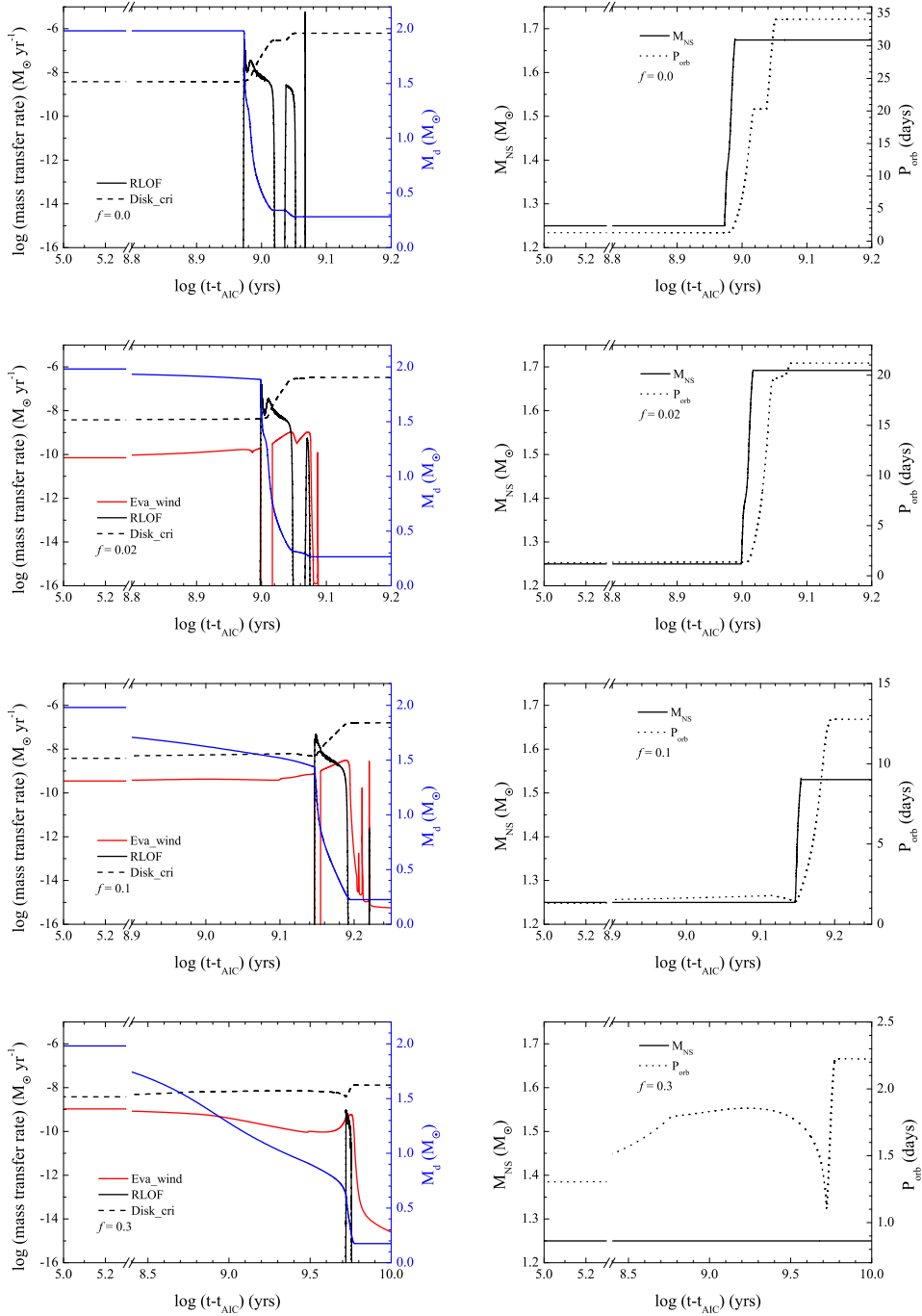

Fig. 6.— Post-AIC evolution of the NS binary with $P_{\rm orb,i} = 1.305$ days and $M_{\rm d,i} = 1.981\ M_\odot$ (corresponding to the blue circle in Fig. 2) with different values of $f$. In the left panel, the red, black solid, black dashed, and blue lines represent the evaporation wind rate, the RLOF mass transfer rate, the critical accretion rate, and the companion mass, respectively. In the right panel the solid and dotted lines represent the NS mass and the orbital period, respectively.

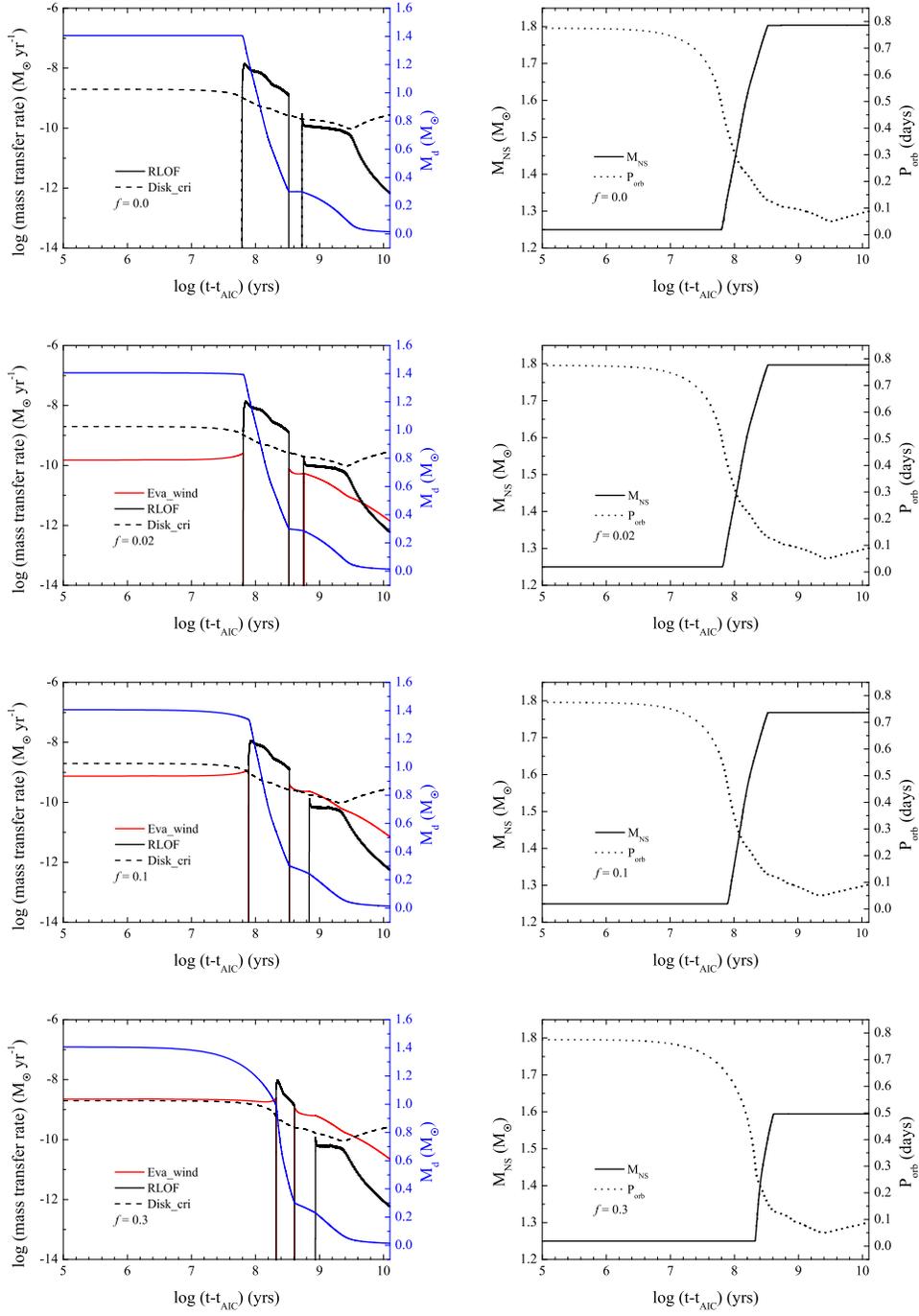

Fig. 7.— Same as Fig. 6, but for the binary with $P_{\rm orb,i} = 0.775$ day and $M_{\rm d,i} = 1.406\,M_\odot$ (corresponding to the red circle in Fig. 2).

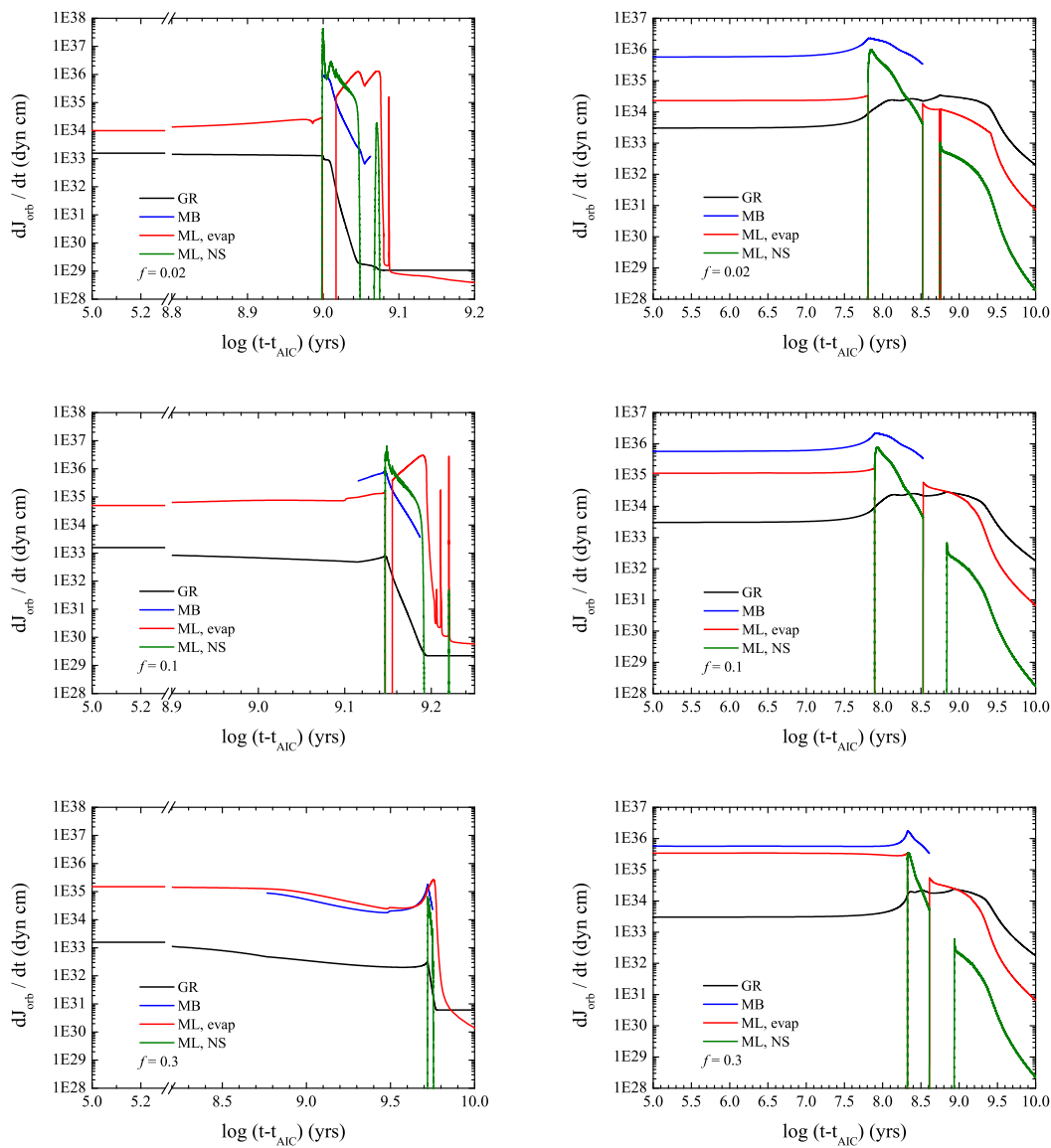

Fig. 8.— Comparison of the AML rates by different mechanisms (black: GR, blue: MB, red: evaporative wind loss from the donor star, green: mass loss from the NS) for the evolution in Figs. 6 (left) and 7 (right), respectively.

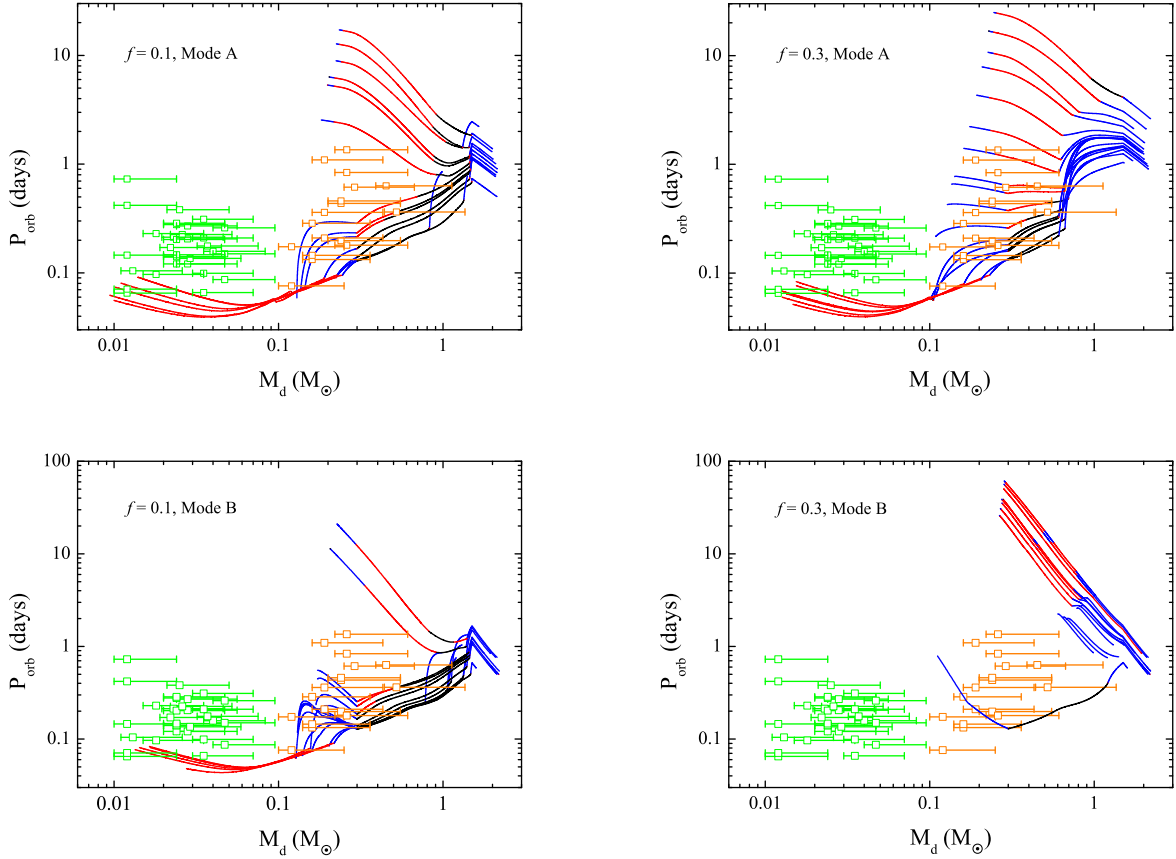

Fig. 9.— Comparison of the calculated evolutionary tracks of the post-AIC binaries with the observations of redbacks (orange squares) and black widows (green squares) in the Galactic disk and globular clusters in the orbital period vs. the donor mass diagram. In the left and right panels $f = 0.1$ and 0.3, respectively. In the upper and lower panels the evaporative winds are assumed to take the specific AM from the surface of the donor star (Mode A) or from the inner Lagrangian point (Mode B), respectively. The blue, black and red lines correspond to RL detachment (with evaporation), stable disk accretion (without evaporation), and unstable disk accretion (with evaporation), respectively. The masses of redbacks and black widows are derived by assuming the orbital inclination angles between 25.8° and 90°.

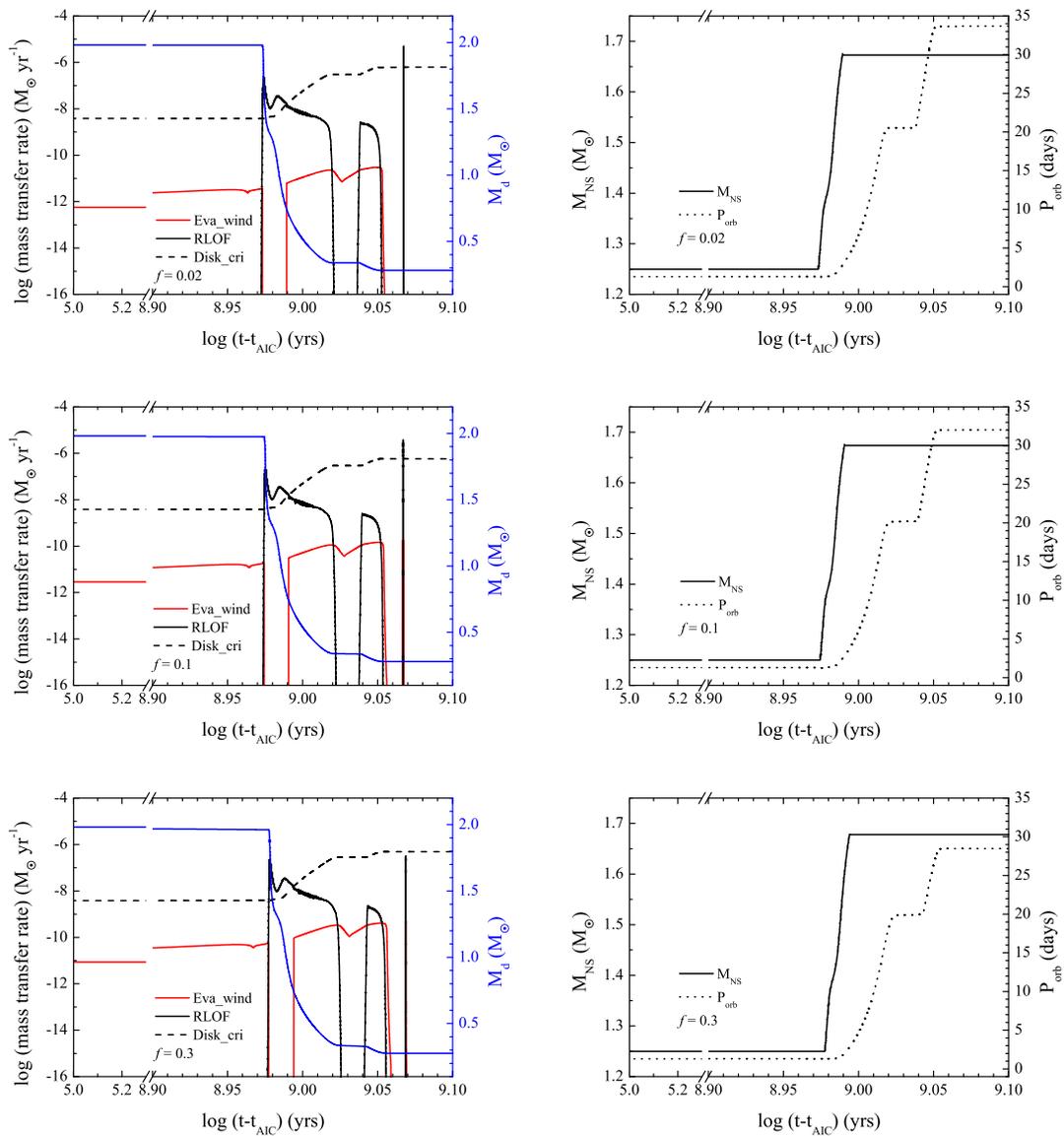

Fig. 10.— Same as Fig. 6, but for $P_{\rm i} = 10$ ms and $B_{\rm surf,i} = 3.0 \times 10^8$ G.

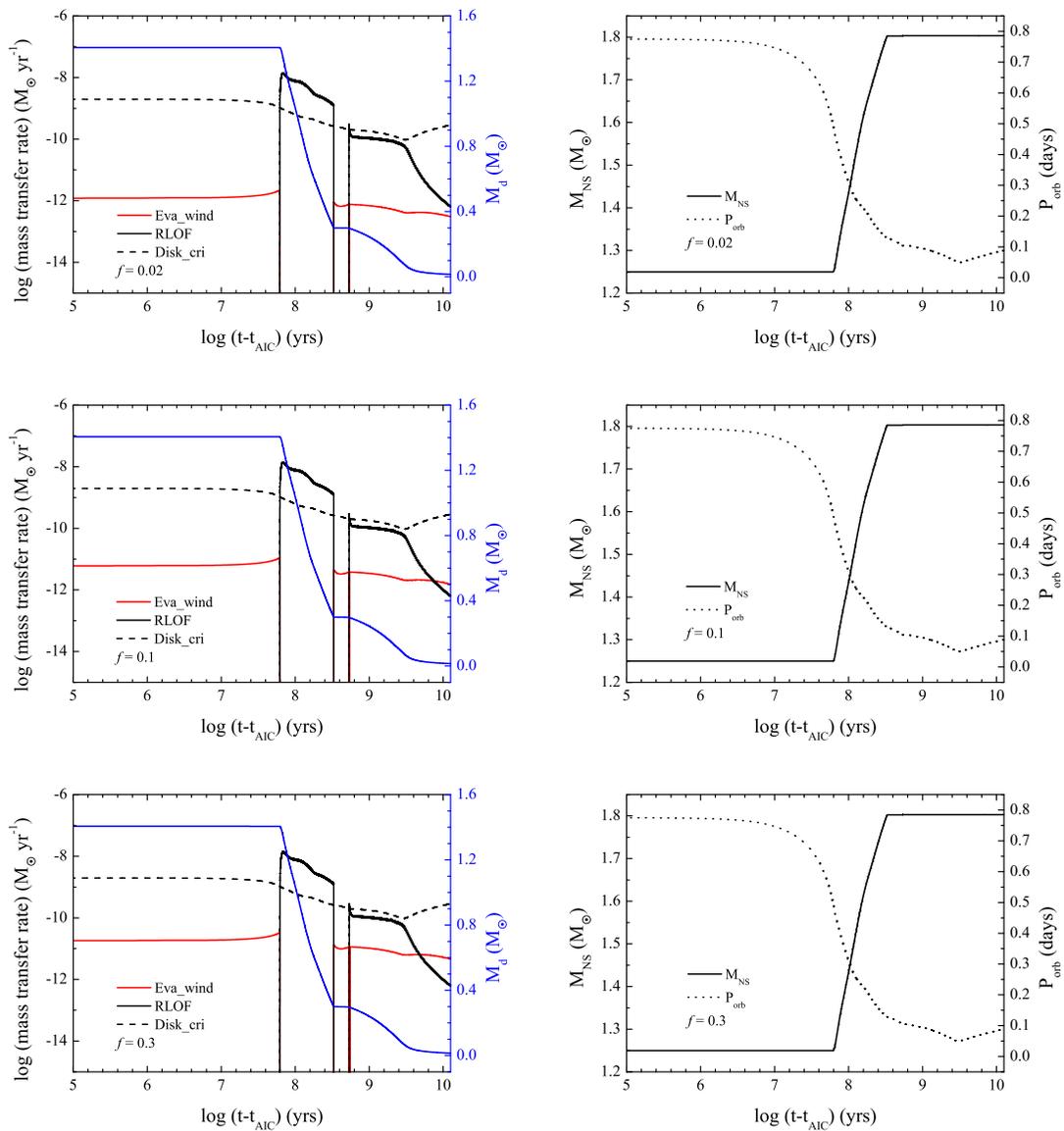

Fig. 11.— Same as Fig. 7, but for $P_{\rm i} = 10$ ms, $B_{\rm surf,i} = 3.0 \times 10^8$ G.

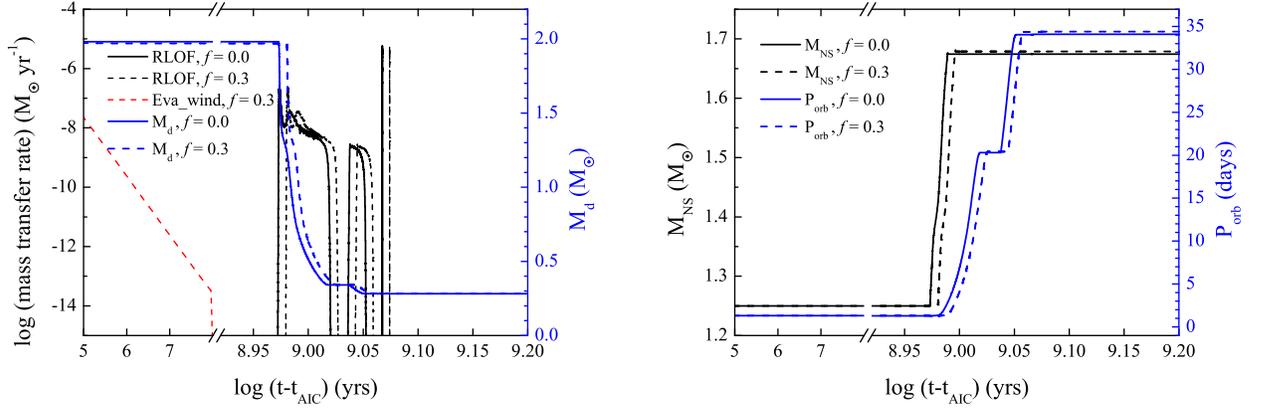

Fig. 12.— Same as Fig. 6 but for $P_i = 3$ ms and $B_{surf,i} = 1.0 \times 10^{12}$ G.

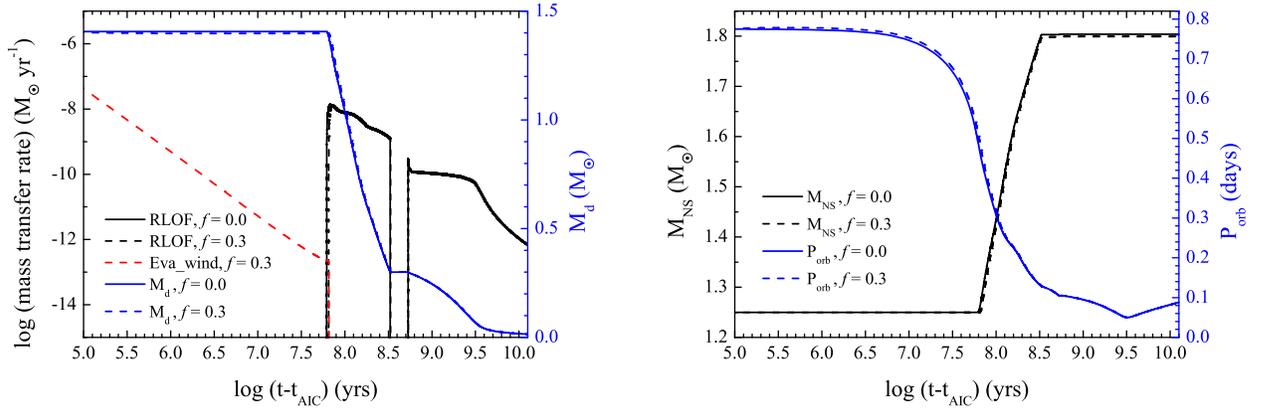

Fig. 13.— Same as Fig. 7 but for $P_i = 3$ ms and $B_{surf,i} = 1.0 \times 10^{12}$ G.